\documentclass{ws-procs975x65}

\begin{document}



\title{QUANTUM GRAVITY AND SPACETIME SYMMETRIES}

\author{RALF LEHNERT}

\address{Center for Theoretical Physics\\
Massachusetts Institute of Technology\\
Cambridge, MA 02139\\
\email{rlehnert@lns.mit.edu}
}


\begin{abstract}
Small violations of spacetime symmetries 
have recently been identified as promising Planck-scale signals. 
This talk reviews how such violations can arise 
in various approaches to quantum gravity, 
how the emergent low-energy effects can be described 
within the framework of relativistic effective field theories, 
how suitable tests can be identified, 
and what sensitivities can be expected in current and near-future experiments.
\end{abstract}
\vskip15pt
\bodymatter

{\bf Introduction.}
One of the most intriguing open questions in current physics research
concerns the structure of spacetime at the Planck length $L_P$. 
While tremendous theoretical efforts 
have been devoted to this subject, 
there is a major obstacle for experimental work: 
the diminutive size of $L_P$. 
A propitious avenue 
to attack this problem 
is provided by ultrahigh-precision tests of symmetries 
that hold exactly in present-day physics 
but might be violated at a more fundamental level. 

In this context, 
violations of Lorentz and CPT invariance 
have recently been found to be promising signatures 
for Planck-length effects \cite{cpt04,proceedings}: 
These symmetries are pillars of established physical laws, 
so that any violation of them 
would indicate qualitatively novel physics. 
In addition, 
Lorentz and CPT tests 
are among the most precise null experiments 
that can be preformed with present or near-future technology. 
Many of these tests 
have Planck reach.
We also mention 
that a number of approaches to underlying physics 
can lead to small Lorentz and CPT breakdown, 
as will be briefly discussed later in this talk. 

Lorentz and CPT symmetry are closely intertwined 
in the CPT theorem, 
which roughly states states 
that a local, unitary, relativistic point-particle quantum field theory 
is CPT invariant. 
One may wonder 
whether CPT and Lorentz invariance 
can be broken independently 
in such a field-theoretical context. 
The answer to this question 
lies in Greenberg's ``anti CPT theorem:'' 
under mild technical assumptions, 
such as unitarity,
CPT violation is always associated with Lorentz breakdown \cite{green}.  
We remark that the opposite, 
namely Lorentz breaking implies CPT violation, 
is false. 
An explicit example for these results 
is given by the Standard-Model Extension, 
which is discussed in the next section.

{\bf Standard-Model Extension.}
For the identification and analysis 
of 
Lorentz and CPT tests, 
a theoretical framework for Lorentz and CPT violation is needed. 
Over the last decade, 
such a framework, 
called the Standard-Model Extension (SME), 
has been developed \cite{sme}. 
This section reviews 
the cornerstones of the SME.

To maintain relative independence 
of the (unknown) underlying physics, 
the SME is {\em constructed} 
to be as general as possible 
while preserving physically desirable features. 
We first use the fact 
that, 
on practical grounds, 
we need a model valid at length scales 
much larger than $L_P$. 
It is then reasonable to assume 
that Lorentz- and CPT-violating effects 
can be described by an effective field theory.\footnote{Effective field theories 
have been tremendously successful in particle and condensed-matter physics. 
The conventional Standard Model itself is usually viewed as an effective field theory, 
so that an effective-field-theory description 
of leading-order Lorentz and CPT violation would seem natural. 
Moreover, discrete backgrounds, 
as might be expected for quantum-gravity effects, 
are known to be compatible with effective field theory, 
at least in solid-state physics.} 
The second basic idea is 
that all of presently established physics 
should be recovered for vanishing Lorentz and CPT violation. 
The desired framework 
is thus a Lagrangian field theory ${\cal L}_{\rm SME}$, 
such that
\begin{equation}
\label{smelagr}
{\cal L}_{\rm SME}={\cal L}_{\rm SM}+{\cal L}_{\rm EH}+\delta {\cal L}\;,
\end{equation}
where ${\cal L}_{\rm SM}$ and ${\cal L}_{\rm EH}$ are the usual Standard-Model
and Einstein--Hilbert Lagrangians, respectively. 
Lorentz- and CPT-breaking effects 
are contained in $\delta {\cal L}$.

For the construction of $\delta {\cal L}$, 
a third ingredient is needed: 
coordinate independence. 
This fundamental principle 
simply states 
that coordinate systems are mathematical tools, 
and as such they should leave unaffected the actual physics. 
It follows that $\delta {\cal L}$ 
must be a coordinate scalar. 
A sample term contained in $\delta {\cal L}$ 
is $\overline{\psi}\gamma_5 b\hspace{-1.6mm}/\psi$, 
where $\psi$ is a fermion field in ${\cal L}_{\rm SM}$ 
and $b^{\mu}$ a small external nondynamical 4-vector 
violating both Lorentz and CPT symmetry. 
In the SME, 
$b^{\mu}$ is a coefficient to be determined by experiment.
Such coefficients are assumed to be generated by underlying physics. 
Some examples are given in the next section.

To date, 
numerous experimental Lorentz and CPT tests 
have been analyzed within the SME \cite{matt05}. 
Studies of cosmic radiation 
have been a particularly popular class of Lorentz tests \cite{dr}. 
The idea is 
that the one-particle dispersion relations 
contain additional Lorentz-breaking terms from $\delta {\cal L}$. 
The resulting modifications 
in particle-reaction thresholds 
would become apparent or more pronounced 
at high energies, 
and they might therefore be observed in cosmic rays. 
An example of such an effect is vacuum \v{C}erenkov radiation \cite{cer}.
If derived within the SME, 
these dispersion-relation corrections are compatible with 
underlying dynamics. 
However, 
the purely kinematical approach 
of postulating modified dispersion relations 
has also been considered \cite{Amelino-Camelia}.

{\bf Sample mechanisms for Lorentz breaking.}
The tensorial coefficients for Lorentz and CPT violation 
contained in the SME
can be generated in a variety of approaches to more fundamental physics.  
This section lists sample theoretical ideas 
that have been developed in this context. 

{\em Spontaneous Lorentz and CPT breakdown in string theory.} ---
From a theoretical
perspective, spontaneous symmetry violation (SSV) is an attractive
mechanism for Lorentz and CPT breaking. 
SSV is well established in
condensed-matter physics, 
and in the electroweak model it is associated with
mass generation. 
The basic idea is 
that a symmetric zero-field configuration 
is not the lowest-energy state. 
Nonzero vacuum expectation values (VEVs) are energetically favored. 
In string field theory, 
it has been demonstrated 
that SSV can trigger VEVs of vector and tensor fields, 
which would then be identified with the Lorentz- and CPT-breaking SME coefficients \cite{ksp}. 

{\em Nontrivial spacetime topology.} --- 
This approach considers the possibility 
that one of the usual three spatial dimensions
is compactified \cite{klink}. 
On observational grounds, 
the compactification radius would be very large. 
Note that
the local structure of flat Minkowski space is preserved. 
The finite size of the compactified dimension 
leads to periodic boundary conditions, 
which implies a discrete momentum spectrum and a Casimir-type vacuum. 
It is then intuitively reasonable 
that such a vacuum possesses a preferred direction
along the compactified dimension.

{\em Cosmologically varying scalars.} --- 
A varying scalar, regardless of the mechanism
driving the spacetime dependence, 
typically implies the breakdown of
translational invariance \cite{spacetimevarying}. 
Since translations and Lorentz transformations are
closely intertwined in the Poincar\'e group, 
it is unsurprising that the translation-symmetry
violation can also affect Lorentz invariance.
Consider, 
for instance, 
a system with
varying coupling $\xi(x)$ 
and two scalar fields $\phi$ and $\Phi$, 
such that the Lagrangian includes a kinetic-type term $\xi(x)\,\partial_{\mu}\phi\,\partial^{\mu}\Phi$. 
A suitable integration by parts generates the term $-(\partial_{\mu}\xi)\,\phi\,\partial^{\mu}\Phi$ 
while leaving unaffected the physics. 
It is apparent 
that the external nondynamical gradient $\partial_{\mu}\xi$ 
can be identified 
with a coefficient of the SME.

{\bf Acknowledgments.}
This work is supported 
by the U.S.\ Department of Energy 
under cooperative research agreement No.\ DE-FG02-05ER41360 
and by the European Commission 
under Grant No.\ MOIF-CT-2005-008687.



\end{document}